\documentclass[twocolumn,secnumarabic,amssymb, nobibnotes, aps, prd]{revtex4-1}

\usepackage {graphicx,epsfig,graphics,color}% Include figure files
\usepackage{dcolumn}% Align table columns on decimal point
\usepackage{bm}% bold math
\usepackage{float}
\usepackage{soul}
\usepackage{changes}

\begin{document}

%\preprint{APS/123-QED}

%\title{\red{\deleted[remark=obsolete]{Direct Observation of magnetic states}Imaging the magnetic states in an actinide %ferromagnet UMn$_2$Ge$_2$} by magnetic force microscopy}% Force line breaks with \\
\title{Imaging the magnetic states in an actinide ferromagnet UMn$_2$Ge$_2$}

\author{Xinzhou Tan}%Lines break automatically or can be forced with \\
\author{Morgann Berg}%
\altaffiliation[Present address: ]{Sandia National Laboratories, Albuquerque, New Mexico 87185, USA}
\author{Alex de Lozanne}
\email{delozanne@physics.utexas.edu}
\affiliation{Department of Physics, University of Texas, Austin, Texas 78712, USA}%
\author{Jeehoon Kim}
\affiliation{Department of Physics, Pohang University of Science and Technology, Pohang 37673, Korea}
\author{R. E. Baumbach}
\altaffiliation[Present address: ]{National High Magnetic Field Laboratory, Florida State University, Tallahassee,
Florida 32310, USA}
\author{E. D. Bauer}
\author{J. D. Thompson}
\author{F. Ronning}
\affiliation{Los Alamos National Laboratory, MPA-CMMS, Los Alamos, New Mexico 87545, USA}%

\date{\today}% It is always \today, today,

\begin{abstract}
We present studies of the magnetic domain structure of UMn$_2$Ge$_2$ single crystals using a home-built low temperature magnetic force microscope. The material has two distinct magnetic ordering temperatures, originating from the Mn and U moments. At room temperature, where the Mn moments dominate, there are flower-like domain patterns similar to those observed in uniaxial ferromagnets. After exposing the sample to a one-tesla magnetic field near 40 K, the evolution of the magnetic domains are imaged through zero-field warming up to 210 K.  Near the ordering temperature of the Uranium moments a clear change in the domain wall motion is observed. The domain size analysis of the flower-like pattern reveals that the domain structure is consistent with a model of branching domains.
\begin{description}
\item[PACS numbers]
May be entered using the \verb+\pacs{#1}+ command.
\end{description}
\end{abstract}

\pacs{Valid PACS appear here}

\maketitle

\section{\label{sec:level1}Introduction}
UMn$_2$Ge$_2$ is a ternary compound belonging to the ThCr$_2$Si$_2$-type family\cite{ThCr2Si2}.This structure is well known for a wide range of intriguing properties such as unconventional superconductivity\cite{super1,super2}. On the other hand, the combination of 5$f$ uranium atoms with 3$d$ transition metal atoms brings up a competition among various energy terms that not only leads to novel transport phenomena like superconductivity and heavy fermion behavior\cite{super3}, but also leads to rich magnetic properties ranging from antiferromagnetism, to ferromagnetism and paramagnetism\cite{A1,A2,A3,A4}. Within this family of compounds UMn$_2$Ge$_2$ is special as both U and Mn ions carry sizable magnetic moments. Magnetization, neutron scattering, and Kerr effect experiments indicate that the Mn ordering temperature $T_c^{Mn}\approx380$ K, while the uranium atoms order below $T_c^{U}\approx150$ K \cite{A150K,B150K,C150K}. Furthermore, recent magnetization measurements under pulsed magnetic fields up to 62 T reveal a strong uniaxial anisotropy with a second order anisotropy constant larger than that of the first order\cite{B}. To date, however, a local and spatially resolved study on the magnetic properties of this compound is still lacking.

\begin{figure*}
\begin{center}
\includegraphics[scale=0.40]{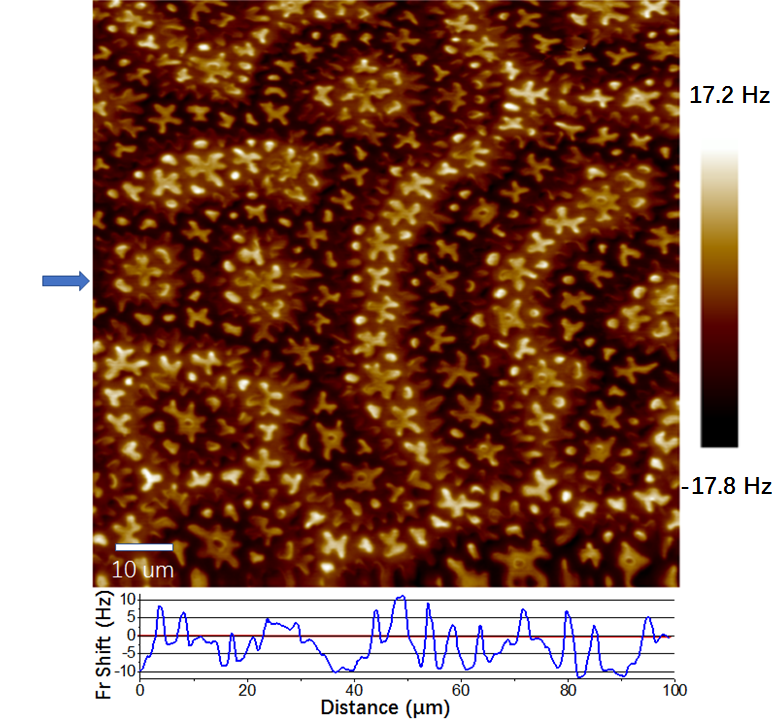}
\caption{A 100 $\times$ 100 $\mu$m$^2$ MFM image taken at room temperature shows typical magnetic domain pattern of UMn$_2$Ge$_2$. The blue array marks the position where the cross section is taken, as shown below the image.
%Color scale corresponds to the resonant frequency shift that is proportional to the magnetic force gradient experienced by the MFM cantilever. The blue array marks the position where the cross section is taken, vertical axis labels the resonant frequency shift of the cantilever.
}
\end{center}
\end{figure*}

Magnetic force microscopy (MFM) has been widely used to map magnetic domain patterns and magnetic phase transitions on various materials\cite{D,C}. In this article, we present magnetic images of a UMn$_2$Ge$_2$ crystal, obtained using a variable-temperature MFM. We have observed the evolution of domain patterns through the magnetic phase transition of the U sublattice and analyzed a flower-like domain with 8-fold symmetry with a possible in-plane anisotropy.

\begin{figure*}
\begin{center}
\includegraphics[scale=0.72]{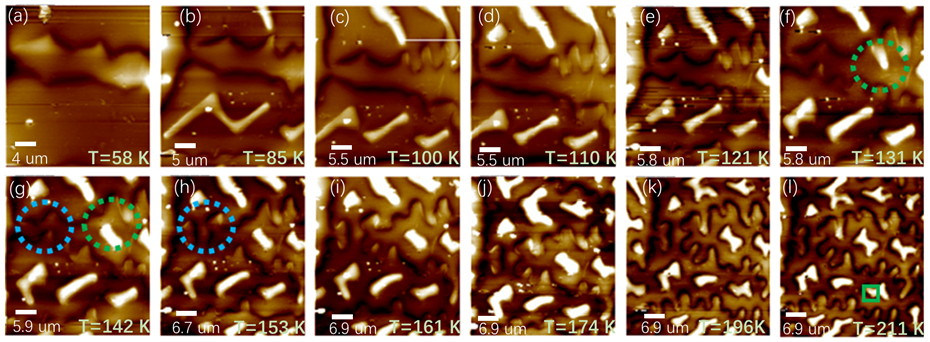}
\caption{\label{fig:wide} Temperature dependent MFM images in the remnant state obtained from  a 1-T field cycle around 40 K. The contrast is adjusted for each image to clearly see the evolution of domain walls. The green and blue dashed circles indicate drastic change of domain walls due to the magnetic phase transition of the uranium moments.}
\end{center}
\end{figure*}
\section{\label{sec:level1}Experimental Details}
%\subsection{\label{sec:level1}Magnetic domain evolution and Uranium atom phase transition}
UMn$_2$Ge$_2$ single crystals are grown from a molten Zn flux as described elsewhere\cite{B}. As confirmed by Laue diffractometry\cite{Morgann}, the sample structure is consistent with the ThCr$_2$Si$_2$ crystal structure and the sample surface is perpendicular to the c axis of the crystal. The rough surface of an as-grown sample brings about many attempts to find flat surfaces suitable for MFM imaging. Details of the experimental setup can be found in ref. \cite{tien}.
\begin{figure}
\begin{center}
\includegraphics[scale=0.5]{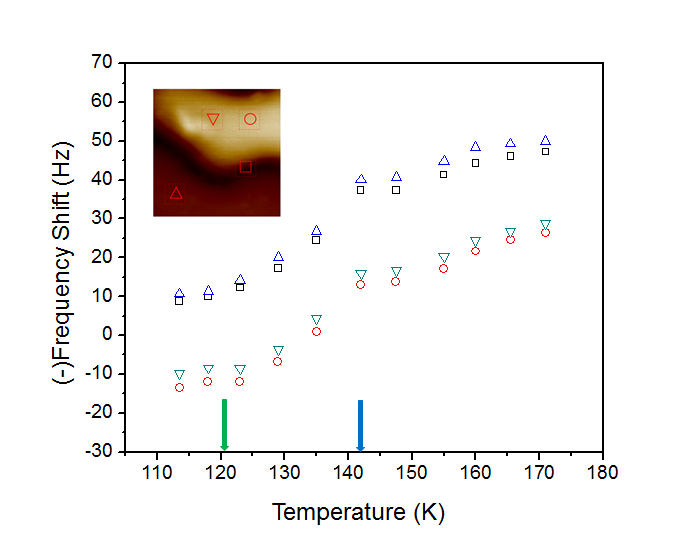}
\caption{Frequency shift versus temperature, taken under zero-field-cooling conditions. The inset shows the area, marked with a green box at $T$ = 211 K in Fig.2, where the measurement is carried out. Four areas are selected to plot the mean frequency shift versus temperature, each with an area around 0.5 $\mu$m by 0.5 $\mu$m, as marked in the inset.
%\A minus sign is added to the vertical axis to show the down trend.
The arrows mark the temperature points right before fast domain motion events discussed in Fig. 2 occur, which correlates well with the turning point brought about by Uranium moment ordering.
}
\end{center}
\end{figure}
To obtain an accurate magnetic mapping of the sample, lift mode is employed to remove spurious signals caused by topography, and the lift height is kept the same to ensure consistency for all data sets. For MFM imaging, the resonant frequency shift is recorded during the lift scan, which is proportional to the magnetic force gradient experienced by the cantilever. Repulsive and attractive forces lead to positive and negative frequency shifts, respectively. Low magnetic moment tips are used because the standard MFM tips give rise to crosstalk between topography and MFM images.

\section{\label{sec:level1}EXPERIMENTAL RESULTS AND DISCUSSION}
Room temperature scans are performed first and one of the results is shown in Fig. 1. The magnetic domains form flower-like patterns along with a maze of curving magnetic stripe domains. The adjacent stripe domains have different contrast, and flower-like domains are in bright contrast relative to their surroundings. Similar contrasts were previously reported in a strong uniaxial ferromagnet Nd$_2$Fe$_{14}$B at room temperature\cite{Hubert1,Hubert2}. By comparing scanning performed at the same area at room temperature and at low temperature down to 77 K separately, it is found that the magnetic domain structure remains unchanged, indicating that the U atom magnetic ordering is not able to alter the existing magnetic domain configuration which presumably is energy-minimized at room temperature. This however is different from Nd$_2$Fe$_{14}$B, for which there is an abrupt magnetic domain pattern reconstruction brought about by a change in the magnetic anisotropy direction from uniaxial axis to an uniaxial cone\cite{Spinchanges}. The stable magnetic pattern strongly implies that there is no significant direction change of the UMn$_2$Ge$_2$ magnetic anisotropy when cooling down to 77K.

The dynamics of the formation and evolution of the domain patterns near the micrometer scale are further investigated.
In particular, the effect of U moment ordering on the evolution of the domains can be directly observed through magnetic imaging. In order to observe the U moment transition through magnetic domains, we drive the domain pattern out of equilibrium by applying a 1-T magnetic field perpendicular to the sample surface at $T \approx$ 40 K, and then we remove the magnetic field and perform zero field warming. The magnetic pattern is initially formed as a metastable state. With increasing temperature, domains start evolving by thermally-driven domain wall motion. As shown in Fig. 2, at $T \approx$ 58 K, the magnetic domains are partially saturated, and more uniform magnetic domains are formed. The domain walls are clearly seen, and the magnetic contrast between different domains was not pronounced compared with domain wall area.

When temperature increases, the wall between big dark domains starts to twist and the major domains grow into each other. The small bright domains form stripes, and then become flower-like. It is worth noting that the big domain walls change rapidly in a certain temperature range (121 K $\sim$ 153 K) below the reported U moment transition temperature, as shown in the dotted circles in Fig. 2 (f)$\sim$ (h). Specifically, between 121 K and 137K there is a quick change of domain wall orientation marked by the green circle, and between 142K and 153K, there is a rapid break of domain walls marked by a blue circle. Since the domain wall of big domains is supposed to extend to the bulk according to the branching domain scheme\cite{branching,book} as will be specified later, these rapid changes are energetically significant compared to changes in bright branching domains. Since similar events are not observed outside of this temperature range, we believe that these fast domain wall motions are associated with the U ordering where U moments start to order ferromagnetically among themselves and against the Mn moments.  %

\begin{figure}
\begin{center}
\includegraphics[scale=0.60]{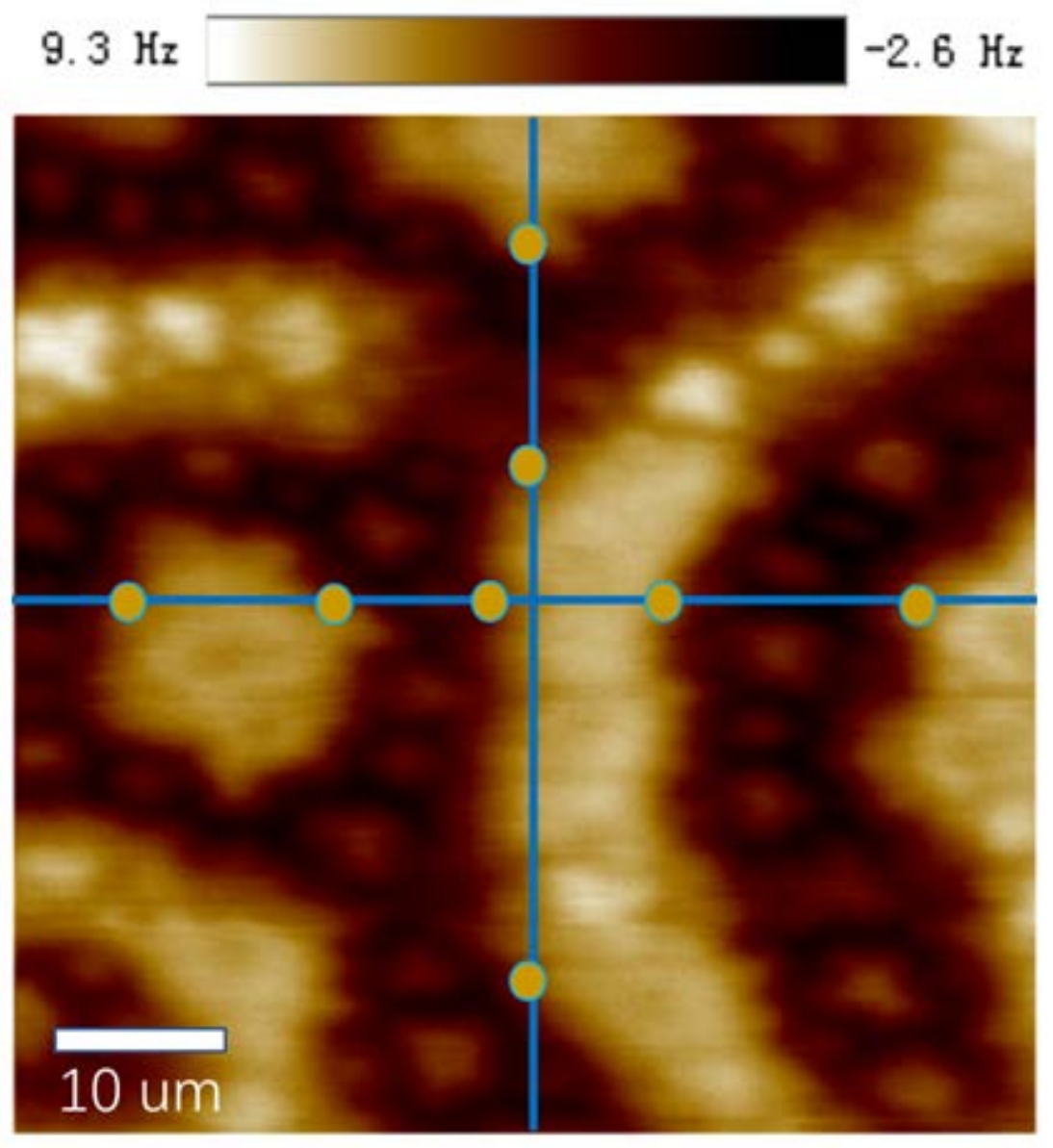}
\caption{Magnetic image with a lift height of about 1 $\mu$m. The small flower-like domains blur out, but the stripe domains underneath persist. The dots mark the intersection points of the test lines for domain width measurement.}
\end{center}
\end{figure}

Further evidence of U moment ordering can be obtained from the frequency shift as a function of temperature, which is approximately proportional to the second derivative of magnetization. Indeed, the frequency shift in an MFM image can be used as a qualitative measure of the local magnetization\cite{Alfred}.
We perform a zero-field cooling experiment and focus on a relatively small area to avoid long scanning times, and the region is labeled in Fig. 2 (l), sitting on the boundary between the big dark domain and a bright domain.

The resonant frequency shift versus temperature on four different locations in the image are shown in Fig. 3. Note that a turning point below 150 K takes place followed with a rapid change before another turning point where the rapid change stops. The temperature points at which the rapid change starts and ends are consistent with temperatures where fast domain wall motions start during the warming sweep in Fig. 2. This gradual, yet prominent U ferromagnetic transition has already been observed by previous investigations\cite{A150K,B150K,C150K}. We believe that the abrupt domain wall motion and frequency shift measurements present the first experimental evidence, on the microscopic scale, of anti-ferromagnetic coupling between U and Mn moments. The anti-ferromagnetic coupling between the U and Mn moments reduces the stray field of the magnetic domains as well as the magnetic dipole-dipole interaction between adjacent magnetic domains. When the anti-ferromagnetic coupling is removed with increasing $T$ toward the U ordering transition temperature, the dipole-dipole interaction potential experiences a rapid increase which enhances effectively domain wall mobility. At high 'magnetic tension' areas such as those labeled by dotted circles in Fig. 2, a drastic increase of the stray field potential overcomes quickly the pinning potential from local defects. This serves as a 'Slingshot' mechanism, driving domain wall motion at a faster rate compared to thermal excitation alone. This result is consistent with what is predicted by first principles calculations\cite{DFT,B}.

It is well known that for strong uniaxial anisotropy magnetic materials such as Nd$_2$Fe$_{14}$B, magnetic domains are governed by a scheme of branching domains in the bulk, whereby large domains inside the sample breakup into smaller domains at the surface\cite{branching,book}. It turns out that UMn$_2$Ge$_2$ has a similar domain structure. Fig. 4 shows an example of room temperature image with lift height around 1 $\mu$m. Following the branching domain scheme, the flower-like domain patterns that fade out at high lift are supposed to be surface domains, and the large stripe domains that are clearly visible should be corresponding to domains penetrating deep into the bulk.
%which is consistent with the model of branching domains, as the volume of the branching flower-like domain is small, %its stray field decays rapidly with distance. On the other hand,
The theoretical model of branching domains uses approximations valid for materials with high uniaxial anisotropy and assumes that spins are oriented either up or down along the easy axis.
The stripe domain width $W_b$, magnetic thin film thickness $D$, and branching surface domain width $W_s$ should follow the following expression derived from equations in ref. \cite{book}:
\begin{equation}
W_b=\sqrt[3]{(4/\pi^2)(0.0185W_s)D^2}
\end{equation}
Our sample thickness is around 0.3 mm. $W_s$ is measured by averaging the width of flower branching domains, which is about 1.8 $\mu$m in our case. This leads to $W_b$$\approx$10.7 $\mu$m.
On the other hand, experimentally, the stripe domain width can be obtained by applying the stereological method \cite{book} on Fig. 4:
\begin{equation}
W_b=\frac{2*Total\\test\\line\\length}{\pi*Number\\of\\intersection}
\end{equation}
For example, in Fig. 4 there are two test lines, with intersections labeled by dots. This measurement gives us a value of $W_b$ $\approx$ 10 $\mu$m. Given the simple nature of this model and the approximations, the good agreement between theory and experiment is better than expected.

\begin{figure}
\begin{center}
\includegraphics[scale=0.35]{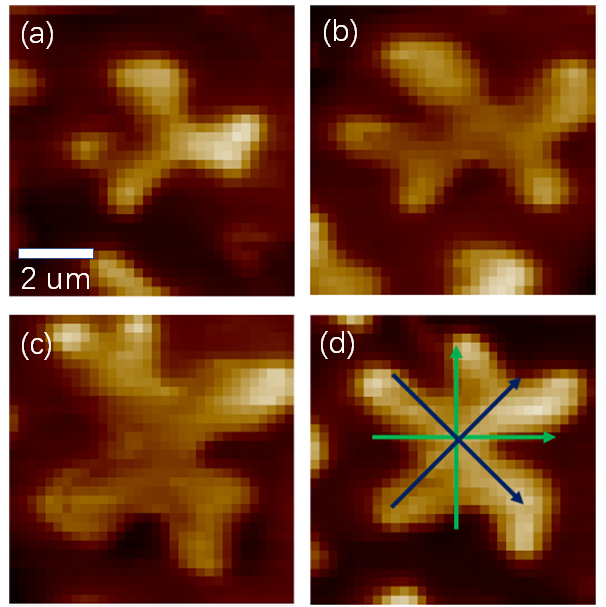}
\caption{Flower patterns at different positions. The shape and orientation of the flower patterns are self-similar. The branching domain walls that enclose the domain align themselves into roughly 4 directions with 45 degree to each other, implying an 8-fold in-plane anisotropy symmetry. The slight mismatch may result from the competition from local closure field energy.
}
\end{center}
\end{figure}

Furthermore, the shape and orientation of these branching flower domains are not at all random, but highly self-similar, as can be seen by comparing different flowers in Fig. 5. Additional evidence comes from the zero-field warming sequence in Fig. 2. At low temperature, the bright branching domains only stretch into two primary directions perpendicular to each other.
As the temperature increases, they break into smaller stripes, and can switch from one direction to the other.
At higher temperature some stripes may choose a direction that is about 45 degrees with respect to the primary direction. This clear directional growth of the branching domains should stem from a requirement to minimize magnetic domain wall energy due to the existence of an in-plane anisotropy, as there is an in-plane moment inside a magnetic domain wall\cite{bloch}. Fig. 5 illustrates flower patterns picked from 4 different locations in Fig. 1. The magnetic domain walls that define the shape of each domain tend to be roughly oriented into 8 directions as marked by arrows in (d).

%By examining each flower-like pattern, one can find 4 directions in which the branch domains initially grow as %appearing in Fig. 2, labeled with green arrow in Fig. 5. Domain walls tend to be parallel with each other when they %become close in order to minimize the dipole-dipole interaction, and this may overcome the major in-plane anisotropy %energy to follow a secondary minima labeled blue in Fig. 5.

By examining each flower-like pattern, one can find 4 dominant directions labeled with blue arrows in Fig. 5.  These directions agree with the orientation of stripe domains seen at low temperatures in Fig. 2.  A less common direction, shown with green arrows, may be the result of dipole-dipole interactions with nearby domain walls, which forces a domain to follow a secondary in-plane minimum.

It is unclear whether this in-plane anisotropy comes from shape anisotropy \cite{book} or is due to spin-orbital coupling induced by a complex anisotropy of the energy surface. For the latter, possible sources for this eight-fold in-plane anisotropy may be the four-fold symmetry of the Mn sublattice as it is projected onto the surface, and the U sublattice which is also four-fold symmetric, but rotated 45 degrees with respect to the Mn lattice.

\vspace{7mm}
\section{\label{sec:level1} Summary}

UMn$_2$Ge$_2$ is an interesting magnetic material with high anisotropy largely due to the combination of U and Mn moments. MFM reveals that the magnetic phase transition of the U moments does not change greatly the magnetic domain landscape, implying that the contribution from the U magnetic moments must closely follow the existing ferromagnetically ordered Mn moments. The 'Slingshot' domain wall motion during a warming sweep provides strong evidence of anti-ferromagnetic coupling between the two moments and the change in resonant frequency shift gives further verification. Moreover, from magnetic domain theory and our MFM experiments, we argue that magnetic domains follow a branching domain scheme, indicating a significant, high $c$-axis anisotropy. Finally, we find the existence of an in-plane anisotropy, which plays an important role in the formation of the delicate flower-like branching domain patterns.

\section{\label{sec:level1} Acknowledgement}
Work at the University of Texas at Austin is supported by NSF grant DMR - 1507874. Work at Pohang University of Science and Technology is supported by the Ministry of Education, Science, and Technology (No. NRF-2017R1A2B4012482). Work at Los Alamos National Laboratory is performed under the auspices of the US Department of Energy, Office of Basic Energy Sciences, Division of Materials Sciences and Engineering.

\end{document}